\documentclass[10pt, conference]{IEEEtran}
\IEEEoverridecommandlockouts
\usepackage{cite}
\usepackage{amsmath,amssymb,amsfonts}
\usepackage{algorithmic}
\usepackage{graphicx}
\usepackage{textcomp}
\usepackage{xcolor}
\usepackage{braket}
\usepackage{array}
\usepackage{tabularx}
\usepackage{booktabs}
\usepackage{makecell}
\usepackage{float}
\usepackage[caption=false,font=footnotesize]{subfig}
\def\BibTeX{{\rm B\kern-.05em{\sc i\kern-.025em b}\kern-.08em
    T\kern-.1667em\lower.7ex\hbox{E}\kern-.125emX}}
\begin{document}

% arXiv preprint: authors in a single line; affiliations in the first-page footnote.
\title{\huge Quantum-Enhanced Classical Communications for DPMZM-based NG-PON2 Systems}

\author{Parvin Kumar, Sourav Banerjee, Neel Kanth Kundu, and Umesh Sharma%
\thanks{Parvin Kumar is with the Bharti School of Telecommunication Technology and Management, Indian Institute of Technology Delhi, New Delhi, India, and the Department of Electronics and Communication Engineering, Krishna Institute of Engineering \& Technology (KIET), Ghaziabad, Uttar Pradesh, India (e-mail: parvin.kaushik@gmail.com).\protect\par
Sourav Banerjee is with the Centre for Applied Research in Electronics, Indian Institute of Technology Delhi, New Delhi, India (e-mail: souravbanerjer9@gmail.com).\protect\par
Neel Kanth Kundu is with the Centre for Applied Research in Electronics and the Bharti School of Telecommunication Technology and Management, Indian Institute of Technology Delhi, New Delhi, India (e-mail: neelkanth@iitd.ac.in).\protect\par
Umesh Sharma is with the Department of Electronics and Communication Engineering, Bennett University, Greater Noida, India (e-mail: umesh356.sharma@gmail.com).}%
\thanks{The work of Neel Kanth Kundu was supported in part by the DST INSPIRE Faculty Fellowship (Reg. No.: IFA22-ENG 344), ANRF Prime Minister Early Career Research Grant (ANRF/ECRG/2024/000324/ENS), and IIT Delhi's New Faculty Seed Grant (MI02855G).}%
\thanks{This paper has been accepted for presentation at IEEE International Conference on Advanced Networks and Telecommunications Systems (ANTS) 2026.}}

\maketitle

\begin{abstract}
This paper presents a framework for integrating quantum detection techniques into a Dual-Parallel Mach–Zehnder Modulator (DPMZM)-based Next Generation Passive Optical Network 2 (NG-PON2) communication system. We present a system model for the DPMZM-based NG-PON2 quantum communication system and characterize its signal-to-noise ratio (SNR), demonstrating its superiority over conventional MZM-based systems. We analyze the performance of various quantum modulation techniques, including binary phase-shift keying, quadrature phase-shift keying, and M-ary quadrature amplitude modulation. A comprehensive analytical framework is developed to evaluate the SNR, BER, and mutual information for both classical and quantum detection schemes within the DPMZM-based NG-PON2 quantum communication system. Our simulation results show that quantum detection significantly outperforms classical detection, and the DPMZM architecture further enhances the performance over the baseline MZM architecture by improving the effective modulation efficiency and receiver sensitivity.

% A Dual-Parallel Mach–Zehnder Modulator (DPMZM) based Next Generation Passive Optical Network 2 (NG-PON2) has been proposed and investigated to improve signal-to-noise ratio, and enhanced spectral efficiency. This paper presents the integration of quantum detection techniques within a DPMZM-based NG-PON2 communication system using modulation schemes including Binary Phase Shift Keying, Quadrature Phase Shift Keying, and M-ary Quadrature Amplitude Modulation. A comprehensive analytical framework is developed to evaluate signal-to-noise ratio, BER, and mutual information for both classical and quantum detection schemes. The results show that quantum detection significantly outperforms classical detection, while the DPMZM architecture further enhances performance through improved effective modulation efficiency and better receiver sensitivity.

\end{abstract}


\section{Introduction}
The Next Generation Passive Optical Network 2 (NG-PON2), standardized under ITU-T G.989, has emerged as a promising technology for future broadband access networks, supporting advanced applications such as 5G/6G backhaul, cloud computing, and the Internet of Things (IoT)\cite{ Nesset2015}, \cite{proakis2008digital}. In addition, the Dual-Parallel Mach--Zehnder Modulator (DPMZM) offers improved linearity, enhanced modulation efficiency, and superior SNR compared with conventional MZM-based systems \cite{patentCN104836625A, Singh2025,9102229, CatalaLahoz2025}. Furthermore, quantum detection techniques improve receiver sensitivity and reduce bit error rate (BER) beyond classical detection limits, particularly in low-photon communication systems \cite{Kondamuri2024}. Recent works have demonstrated that quantum receivers provide superior error performance and enhanced information capacity, highly promising for future high-capacity optical networks \cite{Zhang2023,Pravindra2023,Kondamuri2024}.

Motivated by these developments, this work investigates the integration of quantum detection techniques into DPMZM-based NG-PON2 systems modulation schemes. Prior works on NG-PON2 quantum communication systems have considered MZM-based modulation schemes with on-off keying and BPSK modulation only \cite{Pravindra2023}. Since the operating principles of DPMZM-based NG-PON2 systems differ significantly from conventional MZM-based architectures in terms of optical field superposition, linearity, modulation efficiency, and noise behavior, a dedicated analytical framework is required for accurate modeling and performance evaluation of quantum modulation schemes in DPMZM-based NG-PON2 systems. 

In order to address these challenges, we present a comprehensive system model for a DPMZM-based NG-PON2 architecture, demonstrating its superiority over conventional MZM-based systems in terms of improved signal-to-noise ratio (SNR), enhanced modulation efficiency, and superior linearity against non-linear errors such as third-order intermodulation errors \cite{Kim2011, Torrijos2023, Feng2024}. Next, an analytical framework for various modulation schemes, such as BPSK, QPSK, and M-ary QAM, is formulated in terms of the average photon number $N_s$, the effective modulation index, and the quantum noise statistics \cite{Takeoka2014, Notarnicola2024}. The BER and mutual information are evaluated as functions of $N_s$ under shot-noise-limited and quantum-limited detection conditions, with results benchmarked against classical homodyne bounds and the Helstrom limit and Square Root Measurements (SRM) to quantify the achievable quantum performance gain \cite{Warke2024, Burenkov2021}. The quantum-enhanced detection is compared against classical methods, demonstrating substantial improvements in receiver sensitivity, BER, and mutual information. Finally, a direct architectural comparison between conventional MZM and the proposed DPMZM configuration quantifies the compounded modulation gain and BER advantage delivered by the dual-parallel structure, establishing that the combination of DPMZM transmission and quantum detection yields performance improvements that neither advancement achieves independently \cite{Pravindra2023, Kondamuri2024, Torrijos2023, Feng2024}.

The rest of the paper is organized as follows. The system model and SNR analysis of a DPMZM-based NG-PON2 quantum communication system are discussed in Section \ref{system}. Section \ref{quantum} presents the theory of quantum modulation and detection schemes for enhanced classical communications. The simulation results and the BER comparison with the baseline MZM-based NG-PON2 quantum communication system are presented in Section \ref{sims}. Finally, Section \ref{conc} presents some concluding remarks and future research directions.

% findings of the proposed DPMZM model, comparative BER advantages are provided with respect to the conventional model in Section III. 

\section{System Model} \label{system}

%\begin{figure*}[!t]
%\centering
%\includegraphics[width=\textwidth]{image 1.png}
%\caption{Block diagram of a quantum-based DPMZM-NG-PON2 communication system using adaptive modulation schemes (BPSK/QPSK/M-QAM).}
%\label{fig:quantum_ngpon2}
%\end{figure*}

\begin{figure*}[!t]
\centering
\includegraphics[width=0.8\linewidth]{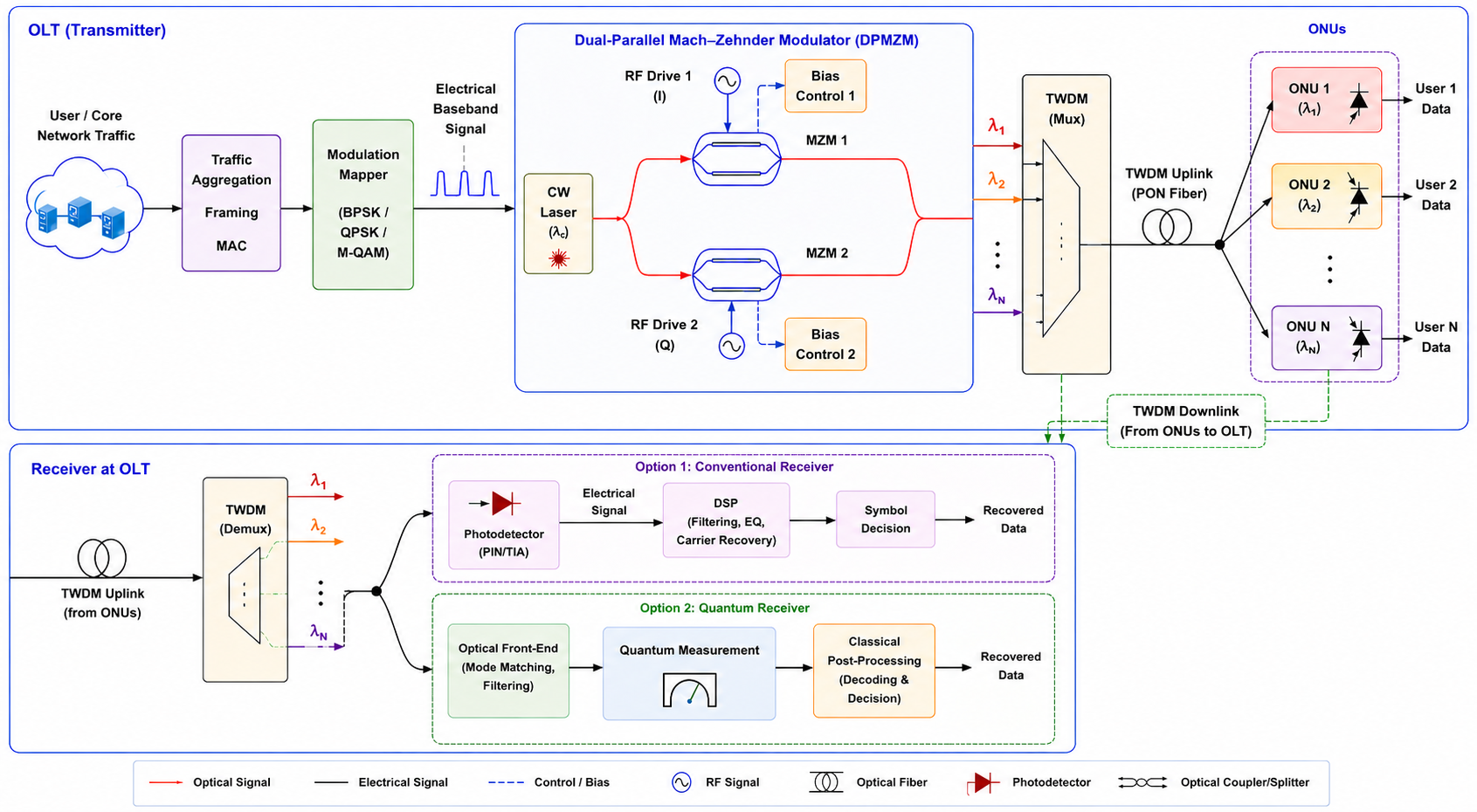}
\caption{Block diagram of a quantum-based NG-PON2 communication system using modulation schemes (BPSK/QPSK/M-QAM)}
\label{fig:quantum_ngpon2}
\end{figure*}

%Fig. 1 illustrates the simplified architecture of a quantum-based NG-PON2 communication system employing a DPMZM for adaptive modulation schemes. At the optical line terminal (OLT), the incoming user or core network traffic is first processed through traffic aggregation, framing, and media access control (MAC) functions. The electrical data stream is then mapped using adaptive modulation formats such as BPSK, QPSK, and M-ary QAM according to system and channel requirements and becomes RF signal. Thereafter, the modulated electrical signal is applied to a continuous-wave (CW) laser source through an external DPMZM, consisting of two parallel MZM branches driven by independent RF signals with bias control. Through Time and Wavelength Division Multiplexing (TWDM), the NG-PON2 architecture enables multiple wavelength channels to serve several Optical Network Units (ONUs) simultaneously. At the receiver side, either a conventional photodetector or a quantum receiver is employed for signal recovery. 

We consider a quantum-enhanced classical communication system. In order to improve the communication performance, the fundamental principles of quantum mechanics are used to encode the classical information using quantum optical states at the transmitter (Alice) and optimally decoded at the receiver (Bob) using optimal quantum measurement strategies.

% According to the fundamental principles of quantum communication, secure information exchange is performed between the transmitter (Alice) and the receiver (Bob), where the transmitted optical states carry information that is optimally decoded at the receiver using advanced quantum measurement strategies.

Fig.\ \ref{fig:quantum_ngpon2} shows the architecture of a quantum-assisted NG-PON2 communication system using DPMZM with modulation schemes such as BPSK, QPSK, and M-QAM. At the Optical Line Terminal (OLT), user data is processed and mapped into modulation symbols before driving a Continuous Wave (CW) laser through the DPMZM structure. The NG-PON2 network employs Time and Wavelength Division Multiplexing (TWDM) technology to support multiple wavelength channels for several Optical Network Units (ONUs) simultaneously. At the receiver, signal detection is performed using either a conventional photodetector or a quantum receiver. In the quantum-enhanced classical communication framework, information is encoded on quantum optical states by the transmitter and efficiently decoded using quantum measurement techniques at the receiver.

The electrical BPSK modulated signal is given as
\begin{equation}
    V(t)=V_m\cos\left(\omega_m t + \pi I(t)\right),
\end{equation}
where $V_m$, $\omega_m$ are amplitude and angular frequency of BPSK modulated signal, $I(t)=\sum_{n} I_n\, g(t-nT)$ is the binary data stream shaped by pulse $g(t)$ over bit duration $T$. The optical output of the DPMZM is given by
\begin{equation}
E(t)=\frac{A_0}{2}e^{j\omega_c t}\left(e^{j\phi_p/2}E_1(t)+e^{-j\phi_p/2}E_2(t)\right)
\end{equation}
where $A_0$ is the launched optical field amplitude, $\omega_c$ is the optical carrier angular frequency, and $\phi_p$ is the parent modulator bias phase. The two sub-MZM branches respond to the quadrature drive signals $V_1(t)=V_m\cos(\omega_m t+\pi I(t))$ and $V_2(t)=V_m\sin(\omega_m t+\pi I(t))$ are defined as
\begin{equation}
E_{i}(t)=\cos\!\left(\frac{\pi V_{i}(t)}{2V_\pi}+\phi_{i}\right), i =1,2 \;.
\end{equation}
Denoting the modulation index as $m=\pi V_m/V_\pi$ with $V_\pi$ being the half-wave voltage of the DP-MZM modulator, the effective index of the dual-parallel structure is given by
\begin{equation}{\label{eq:m_eff}}
m_\text{eff}=\frac{1}{2}\left(m_1+m_2+2\sqrt{m_1 m_2}\cos\phi_p\right)
\end{equation}
which reduces to $m_\text{eff}=2m$ under equal branch drive ($m_1=m_2=m$) and ideal constructive bias ($\phi_p=0$). The output field then simplifies to
\begin{equation}
E(t)\approx A_0 e^{j\omega_c t}\left(1+m_\text{eff}\cos(\omega_m t+\pi I(t))\right)
\end{equation}

At the receiver, direct detection yields the photocurrent as
\begin{equation}
\begin{split}
I(t) &=
\mathcal{R}
A_0^2\,M_a\,G\,G_e\,
10^{-\alpha_1 l/10}
\\
&\quad \times
\Big(
1+m_{\mathrm{eff}}^{2}
+2m_{\mathrm{eff}}
\cos\!\big(\omega_m t+\pi I(t)\big)
\Big)
+n(t)
\end{split}
\end{equation}
where $l$ denotes the fiber length (in Km), $\alpha_1$ is the attenuation coefficient (in dB/Km), $M_a$ is the avalanche gain, $G$ is the optical gain, and $G_e$  is the electrical amplifier gain. Furthermore,  $\mathcal{R} = \frac{\eta e}{h\nu}$ is the photodetector responsivity, where $\eta$ is the quantum efficiency of the photodetector, $\nu$ the optical carrier frequency, and $n(t)$ the additive noise. The mean-square signal current is given by
\begin{equation}
\langle I^2(t)\rangle=\left(
\frac{\eta e}{h\nu}
\right)^2 A_0^4\left(M_a G G_e\right.10^{-\alpha_1 l/10})^2 \left(\frac{m_\text{eff}^2}{2}\right)
\end{equation}
The received optical power due to fiber attenuation is given by
\begin{equation}
P_0 = A_0^2 \cdot 10^{-\alpha_1 l/10}
\end{equation}
Therefore, the mean-square signal current can be expressed as
\begin{equation}
\left\langle I^2(t) \right\rangle
=
\frac{m_{\mathrm{eff}}^2}{2}
\left(
\frac{\eta e}{h\nu}
\right)^2
(G\,G_e\,M_a)^2
P_0^2
\end{equation}
In the shot-noise limited regime, the mean-square shot noise at the receiver is given by
\begin{equation}
\left\langle i_{\mathrm{shot,RX}}^{2} \right\rangle
=
2eI_pB_eM_a^{2+x}
\end{equation}
where $e$ is electron charge, $B_e$ is electrical bandwidth is equivalent to $B/2$, $B$ is bit rate, $x$ is the excess noise factor and $I_p$ is the mean photocurrent given by $$ I_p =
\frac{\eta e}{h\nu}\,G\,G_e\,M_a\,P_0.$$
Hence, 
\begin{equation}
\left\langle i_{\mathrm{shot,RX}}^{2} \right\rangle
=
\frac{2e^{2}\eta  G  G_e  B_e  M_a^{3+x} P_0}
{h\nu}
\end{equation}
With shot noise as the dominant noise source and received power as $P_0=N_sh\nu B$, where $N_s$ is the average received photon number per symbol, the SNR of the DPMZM-based NG-PON2 system is given by
% \begin{equation}
% \mathrm{\text{SNR}_{\text{DPMZM}}}
% =
% \frac{\left\langle I^{2}(t)\right\rangle}
% {\left\langle i_{\mathrm{shot,RX}}^{2}\right\rangle}
% \end{equation}
\begin{align}{\label{eq:SNR_DPMZM}}
\mathrm{\text{SNR}_{\text{DPMZM}}}
&= 
\frac{\left\langle I^{2}(t)\right\rangle}
{\left\langle i_{\mathrm{shot,RX}}^{2}\right\rangle}  =
\frac{
\frac{m_{\mathrm{eff}}^{2}}{2}
\left(\frac{\eta e}{h\nu}\right)^{2}
G^{2}G_{e}^{2}M_{a}^{2}P_{0}^{2}
}{
\frac{2e^{2}\eta G G_{e} B_{e} M_{a}^{3+x} P_{0}}
{h\nu}
} \nonumber 
\\[6pt]
% &=
% \frac{
% m_{\mathrm{eff}}^{2}\eta^{2}e^{2}G^{2}G_{e}^{2}M_{a}^{2}P_{0}^{2}
% }{
% 2(h\nu)^{2}
% }
% \cdot
% \frac{h\nu}
% {
% 2e^{2}\eta G G_{e} B_{e} M_{a}^{3+x} P_{0}
% }
% \\[6pt]
&=
\frac{
m_{\mathrm{eff}}^{2}\eta G G_{e} P_{0}
}{
4h\nu B_{e} M_{a}^{1+x}
}.
\end{align}

The SNR of the baseline MZM-based NG-PON2 system is given by \cite{Pravindra2023}
\begin{equation}{\label{eq:SNR_MZM}}
\mathrm{\text{SNR}_{\text{MZM}}}
=
\frac{m^2 G G_e \eta P_0}
{4 B_e h \nu M_a}
\end{equation}
From the theoretical SNR expressions of the proposed DPMZM and baseline MZM presented in \eqref{eq:SNR_DPMZM} and \eqref{eq:SNR_MZM}, respectively, with $m_\text{eff}=2m$, it is evident that the proposed DPMZM architecture provides an SNR enhancement of approximately $6\,\mathrm{dB}$ compared to the conventional MZM-based system. This substantial improvement demonstrates the effectiveness of the proposed scheme and makes it a promising candidate for NG-PON2 applications requiring high-performance and reliable communication links, as it enhances receiver sensitivity and supports more reliable optical transmission.

\section{Quantum Modulation and Detection Schemes} \label{quantum}
The quantum detection framework for coherent-state modulation is unified across BPSK, QPSK,
and M-QAM through the common language of Hilbert space geometry, Gram matrix analysis, and Square Root Measurement (SRM) theory. Each modulation scheme admits a distinct signal
representation, overlap structure, and optimal receiver strategy, yet all share the same
fundamental trade-off between photon number and detection reliability.

\subsection{Signal Representations}

In BPSK, the two equiprobable symbols are encoded as antipodal coherent states
\begin{equation}
    |\psi_0\rangle = |\alpha\rangle, \qquad |\psi_1\rangle = |-\alpha\rangle,
\end{equation}
where $ N_s = |\alpha|^2$ is the average number of photons.
QPSK extends this to four phase-symmetric states given by
\begin{equation}
    |\psi_k\rangle = \left|\sqrt{N_s}\,e^{j2\pi k/4}\right\rangle, \qquad k = 0,1,2,3,
\end{equation}
which are uniformly distributed on a circle of radius $\sqrt{N_s}$ in phase space. For square $M$-QAM
with $M = L^2$, the constellation is a rectangular lattice of coherent states
\begin{equation}
    |\psi_m\rangle = |\Delta(u_m + jv_m)\rangle
\end{equation}
where the spacing parameter
\[
 \Delta = \sqrt{\frac{3N_s}{2(L^2-1)}}
\]
normalises the average photon number to $N_s$ across all $M$ constellation points \cite{Gianfranco2015}.

\subsection{State Overlap and Gram Matrix}

The distinguishability of quantum states is governed entirely by their pairwise inner
products. The overlap between two coherent states is given by
\begin{equation}
\langle \psi_p | \psi_q \rangle = \exp\left(-\frac{1}{2}|\alpha_p - \alpha_q|^2 + j\,\mathrm{Im}(\alpha_p^*\alpha_q)\right)
\end{equation}

For BPSK, the overlap takes the closed form 
\begin{equation}
    \langle\psi_0|\psi_1\rangle = e^{-2N_s},
\end{equation}
which decays exponentially with photon number and quantifies the irreducible quantum
indistinguishability of the two signals. For QPSK constellation, the pairwise overlaps between states
$p$ and $q$ are given by
\begin{equation}
    G_{pq} = \langle\psi_p|\psi_q\rangle= \exp\!\left[-N_s\!\left(1 - e^{j2\pi(q-p)/4}\right)\right],
\end{equation}
which are assembled into a $4\times 4$ Gram matrix $\mathbf{G}$. For the M-QAM modulation, the Gram matrix is $M\times M$ dimensional whose elements admit
\begin{equation}
\begin{split}
G_{mn}
&= \langle\psi_m|\psi_n\rangle \\
&= \exp\!\left[
-\frac{\Delta^2}{2}\bigl((u_m-u_n)^2+(v_m-v_n)^2\bigr)
\right. \\
&\qquad\left.
+\, j\Delta^2(u_m v_n - v_m u_n)
\right]
\end{split}
\end{equation}
which simultaneously encodes both amplitude separation and phase rotation between
constellation points. As the constellation order increases at fixed $N_s$, the minimum
inter-state distance decreases, the off-diagonal elements of $\mathbf{G}$ grow, and
quantum state discrimination becomes fundamentally harder.

\subsection{Optimal Receivers}

% For BPSK, the optimal strategy is the Helstrom minimum-error measurement, which exploits
% the two-dimensional geometry of the signal subspace directly. \textbf{\color{blue}[Note that for $M>2$, the optimal receiver is semidefinite programming based, but it is difficult to obtain due to high computational complexity; therefore, in practice we use SRM, which is a pretty good measurement]} For QPSK and M-QAM, the
% optimal receiver is the Square Root Measurement (SRM) constructed from the Gram matrix.
% The SRM positive operator-valued measure (POVM) is defined via the matrix square root
% $\mathbf{G}^{1/2}$, and the conditional correct-detection probabilities are given by \textbf{\color{blue}[You can present a generalized SRM framework for QPSK and QAM (No need to use two equations for the conditional probabilities of QPSK and QAM)].}
% \begin{equation}
%     p_c^{\mathrm{QPSK}}(p|q) = \left|\left(\mathbf{G}^{1/2}\right)_{pq}\right|^2 \qquad,
%     \qquad
%     \end{equation}
%  \begin{equation}   
%     p_c{\mathrm{M-QAM}}(m|n) = \left|\left(\mathbf{G}^{1/2}\right)_{mn}\right|^2 \qquad
% \end{equation}
% The BPSK case reduces to a binary decision governed directly by the Helstrom bound,
% without requiring eigendecomposition.
For BPSK modulation, the optimal detection strategy is the Helstrom minimum-error measurement, which directly exploits the two-dimensional geometry of the signal subspace. For binary coherent-state discrimination, the Helstrom receiver achieves the minimum achievable error probability and does not require eigendecomposition of the Gram matrix.

For modulation schemes with $M>2$ constellation points, obtaining the true optimal quantum receiver generally requires solving a semidefinite programming (SDP) problem. Although SDP-based receivers are theoretically optimal, their computational complexity grows rapidly with constellation size, making them impractical for most applications. Consequently, the Square Root Measurement (SRM) is often employed as a practical alternative due to its relatively low computational complexity and its performance, which is typically very close to that of the SDP-derived optimal receiver. Furthermore, it has been shown that SRM is optimal for pure states with a geometrically uniform symmetry (GUS) constellation, which is valid for the QPSK modulation \cite{dalla2015optimality}.

Accordingly, for QPSK and $M$-QAM constellations, we employ a generalized SRM framework. Let $\mathbf{G}$ denote the Gram matrix of the coherent-state constellation. The SRM measurement operators are constructed from the matrix square root $\mathbf{G}^{1/2}$, and the corresponding conditional detection probabilities are given by
\begin{equation}
    p_c(i|j)
    =
    \left|
    \left(\mathbf{G}^{1/2}\right)_{ij}
    \right|^2,
\end{equation}
where $i,j \in \{1,2,\ldots,M\}$ index the transmitted and detected constellation symbols, respectively. This formulation applies uniformly to both QPSK and higher-order $M$-QAM constellations.
\subsection{Minimum Error Probability}

The minimum achievable error probabilities under optimal quantum detection are given by 
% \begin{equation}
% P_e^{\mathrm{BPSK-Homodyne}} = Q\left(2\sqrt{N_s}\right)
% \end{equation}
\begin{equation}
P_e^{\text{Helstrom}} = 1 - \frac{1}{M}\max_{\{\Pi_i\}} \sum_{i=1}^{M} \text{Tr}(\Pi_i \rho_i)
\label{eq:helstrom_general}
\end{equation}
\[\text{subject to} \quad \Pi_i \geq 0, \quad \sum_{i=1}^{M} \Pi_i = \mathbb{I} \]

\begin{equation}
    P_e^{\mathrm{BPSK-Helstrom}}
    = \frac{1}{2}\!\left(1 - \sqrt{1 - e^{-4N_s}}\right),
\end{equation}
\begin{equation}
    P_e^{\mathrm{QPSK}}
    = 1 - \frac{1}{16}\!\left(\sum_{i=1}^{4}\sqrt{\lambda_i}\right)^{\!2},
\end{equation}
\begin{equation}
    P_e^{\mathrm{QAM}}
    = 1 - \frac{1}{M^2}\!\left(\sum_{i=1}^{M}\sqrt{\lambda_i}\right)^{\!2},
\end{equation}
where $\{\lambda_i\}$ denotes the non-negative eigenvalues of the respective Gram matrix.
All three expressions share the same structural dependence on $N_s$: as photon number
increases, the Gram matrix approaches the identity, its eigenvalues converge to unity,
and the error probability tends to zero. The BPSK bound is exact and analytic; the QPSK
and M-QAM bounds require numerical eigendecomposition of $\mathbf{G}$, whose dimension
grows as $M$ increases.

\subsection{Mutual Information}

The achievable mutual information under SRM detection quantifies the maximum rate at
which transmitted symbols can be reliably inferred from measurement outcomes. For BPSK,
the symmetric binary structure yields the compact expression
\begin{equation}
    I_{\mathrm{BPSK}} = 1 - H_2\!\left(P_e^{\mathrm{BPSK}}\right),
\end{equation}
where $H_2(p) = -p\log_2 p - (1-p)\log_2(1-p)$ is the binary entropy function.
For QPSK and M-QAM, the mutual information must be computed from the full transition
probability matrix derived from $\mathbf{G}^{1/2}$ as  \cite{Kato1999}
\begin{equation}
    I_{\mathrm{QPSK}}
    = \frac{1}{4}\sum_{k}\sum_{y}
      p_c(y|k)\log_2\frac{4\,p_c(y|k)}{\displaystyle\sum_{j}p_c(y|j)},
\end{equation}
\begin{equation}
    I_{\mathrm{QAM}}
    = \frac{1}{M}\sum_{k}\sum_{y}
      p_c(y|k)\log_2\frac{M\,p_c(y|k)}{\displaystyle\sum_{j}p_c(y|j)}.
\end{equation}
In the limit $N_s \to \infty$, all three expressions saturate their respective maxima
of $1$, $2$, and $\log_2 M$ bits per symbol. At low photon numbers, higher-order
modulation suffers a steeper collapse in mutual information due to the simultaneous
increase in the number of overlapping states and the reduction in inter-state separation,
making quantum detection particularly critical in this regime.

\

\section{Numerical Results} \label{sims}

\begin{figure}[!t]
\centering
\includegraphics[width=1\columnwidth]{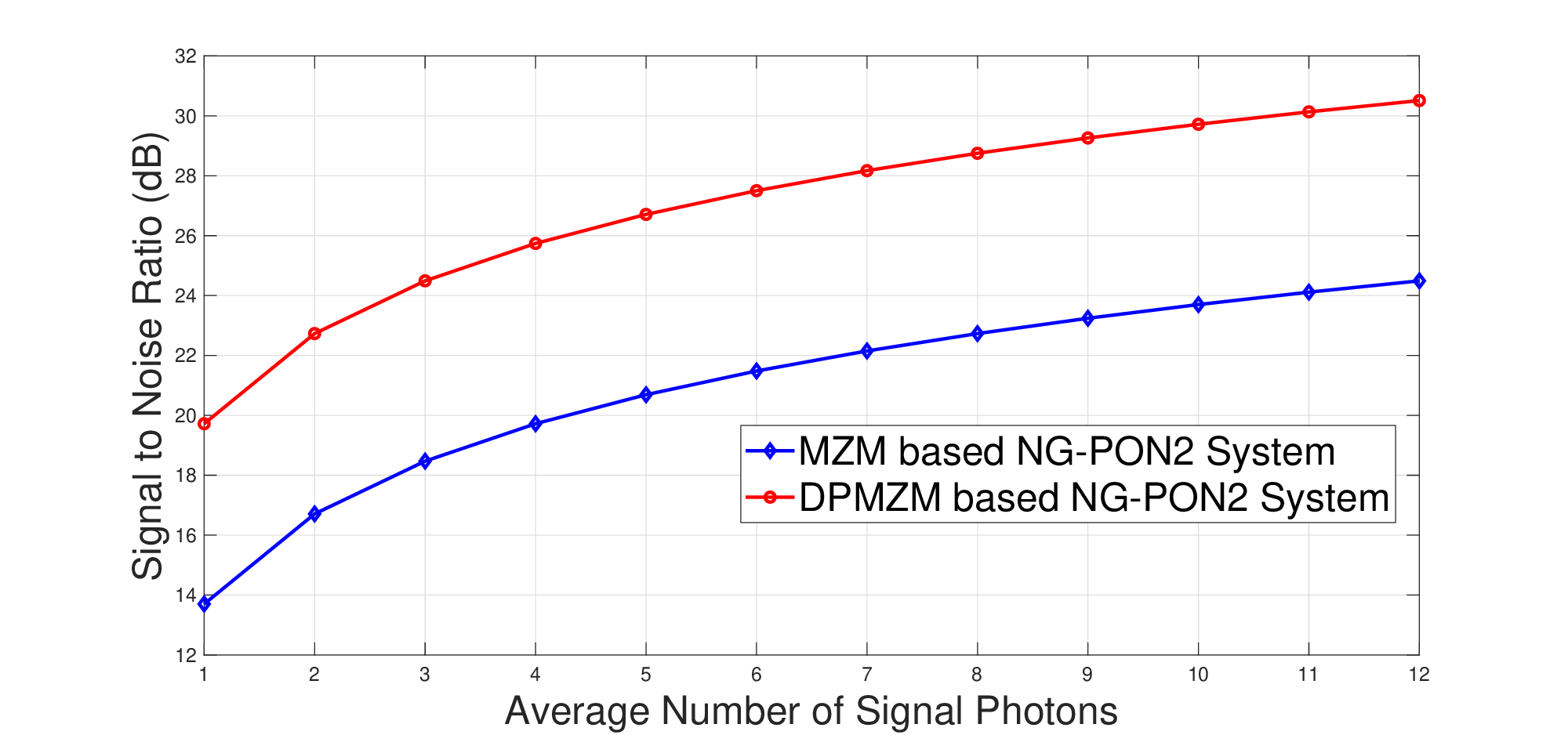}
\caption{SNR versus average number of signal photons \(N_s\) for MZM-based and DPMZM-based NG-PON2 communication systems.}
\label{fig:snr_ns_compare}
\end{figure}

To compare the performance of the proposed DPMZM model with conventional MZM architecture, the parameters are considered as follows: modulation index $m=0.25$, efficiency $\eta=0.75$, an optical fiber length of $l=20$ km with $\alpha=0.2$ dB/km attenuation, the excess noise factor of $x=1$ for avalanche photodetector and the photodetector responsivity $\mathcal{R}=0.954$ A/W. Fig.~\ref{fig:snr_ns_compare} compares the SNR of the DPMZM- and MZM-based architectures as a function of the average received signal photons \(N_S \). The SNR increases with \(N_S\) for both systems, while the proposed DPMZM architecture maintains an approximately \(6\)-dB advantage across the considered range due to its higher effective modulation index. This improvement enhances receiver sensitivity and link margin, demonstrating the potential of the proposed DPMZM-based NG-PON2 system for low-power, quantum-enhanced optical access networks.
\begin{figure}[t]
\centering

\subfloat[BPSK]{%
\includegraphics[width=0.5\textwidth]{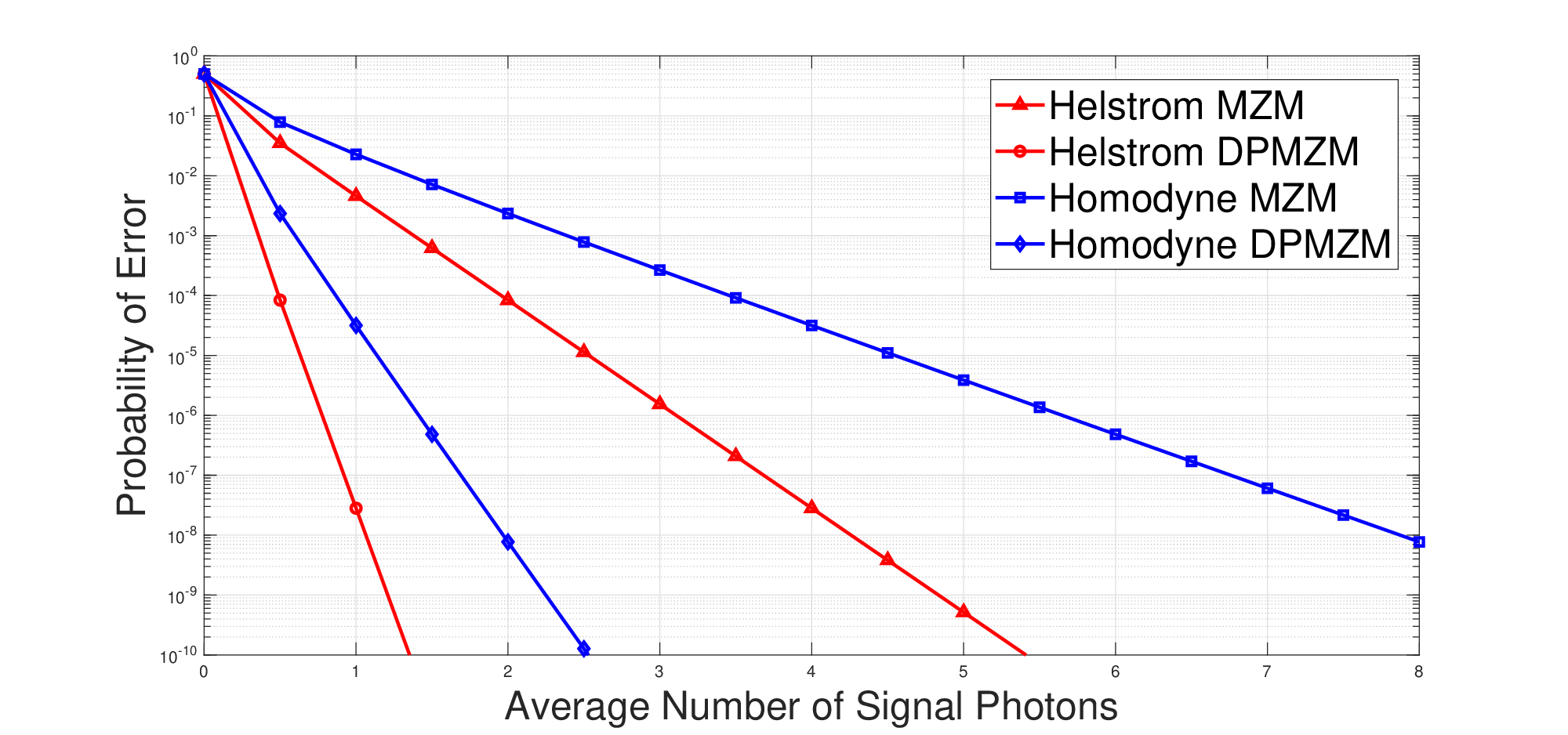}
\label{fig:fig2a}
}
\hfill
\subfloat[QPSK]{%
\includegraphics[width=0.5\textwidth]{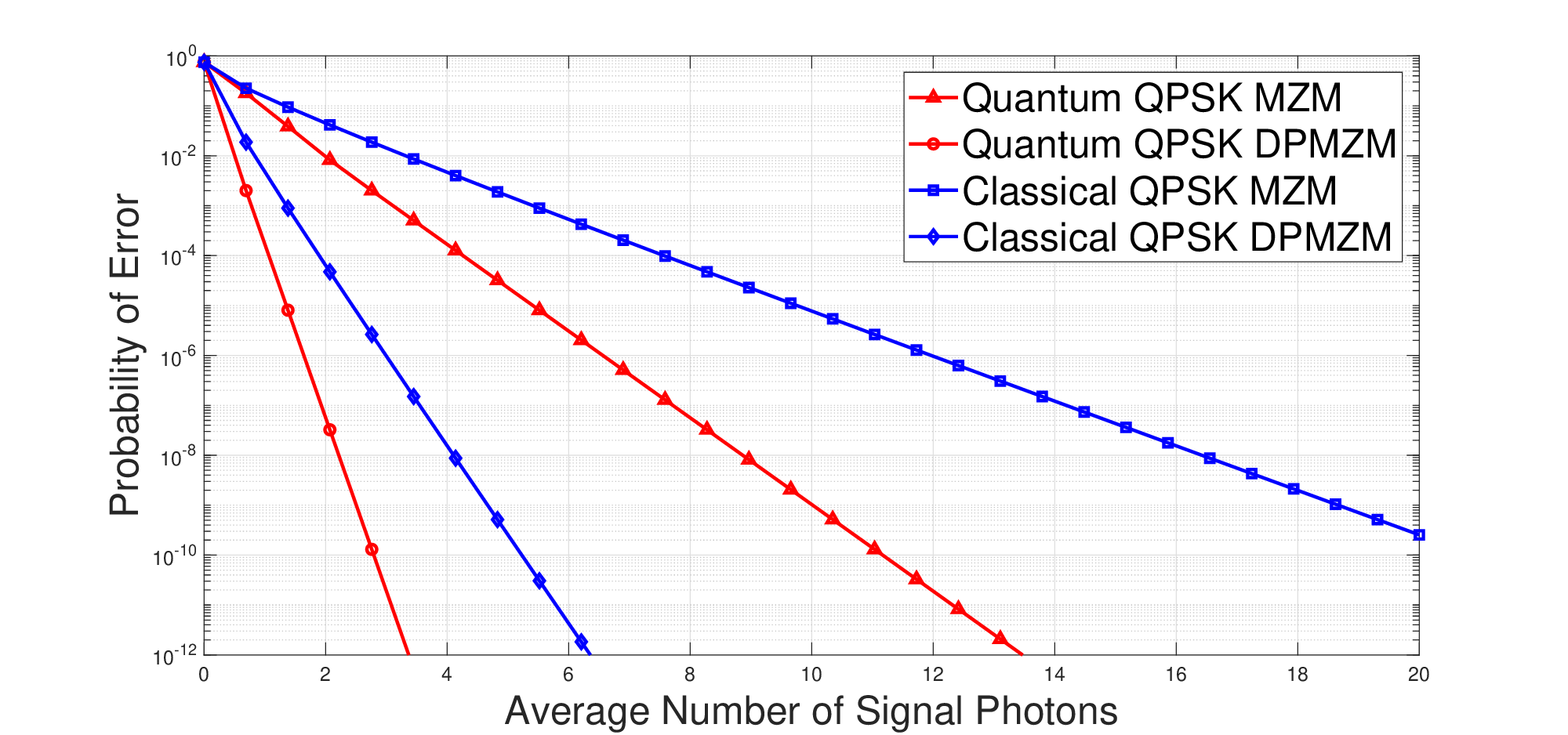}
\label{fig:fig2b}
}

\vspace{0.3cm}

\subfloat[16-QAM]{%
\includegraphics[width=0.5\textwidth]{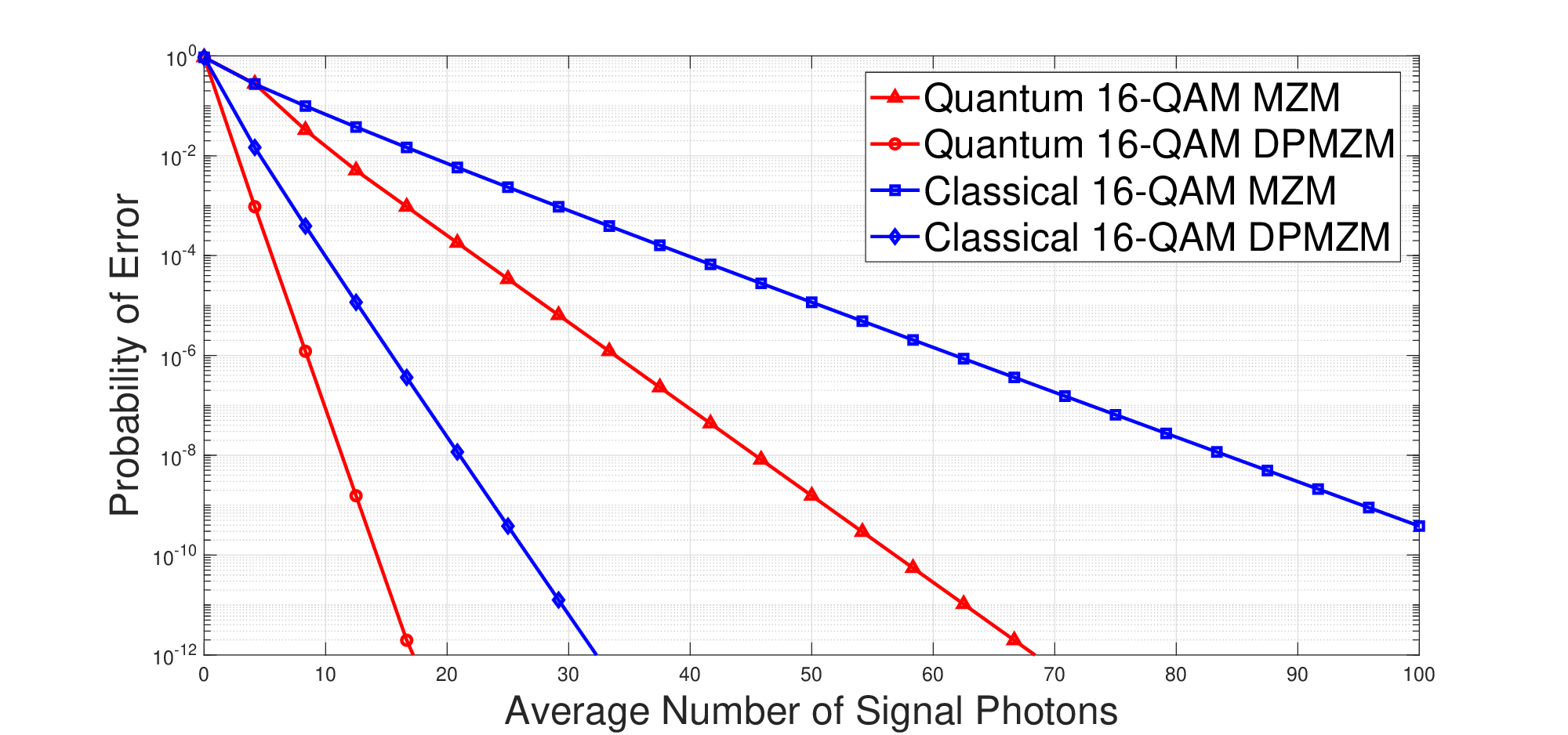}
\label{fig:fig2c}
}

\caption{BER performance versus average signal photon number for BPSK, QPSK, and 16-QAM schemes.}
\label{fig:fig2}

\end{figure}

Next, the performance of quantum and classical detection schemes for BPSK, QPSK, and $16$-QAM modulation formats is studied in terms of probability of error as a function of the average number of signal photons per symbol $N_s$. The results are illustrated in Fig. \ref{fig:fig2}. As shown in Fig. \ref{fig:fig2a}, for BPSK modulation, at \(N_s = 1\), the proposed quantum-based DPMZM system achieves nearly five orders of BER improvement compared with quantum-based MZM and about three orders of improvement compared with classical detection using DPMZM. The Helstrom receiver achieves the minimum probability of error, representing the ultimate quantum limit. The homodyne receiver closely follows this bound.
For QPSK and $16$-QAM modulation, the quantum receiver is implemented using SRM. As shown in Fig.~\ref{fig:fig2b} for QPSK modulation, at \(N_s = 5\), the proposed quantum-based DPMZM system achieves nearly nine orders of BER improvement compared with quantum-based MZM and about four orders of improvement compared with classical detection using DPMZM. This improvement is attributed to the optimal discrimination of non-orthogonal coherent states using SRM, which effectively exploits the quantum structure of the signal space. The results highlight that quantum detection provides considerable gains even for QPSK.
\begin{figure}[h]
\centering

\subfloat[BPSK]{%
\includegraphics[width=0.5\textwidth]{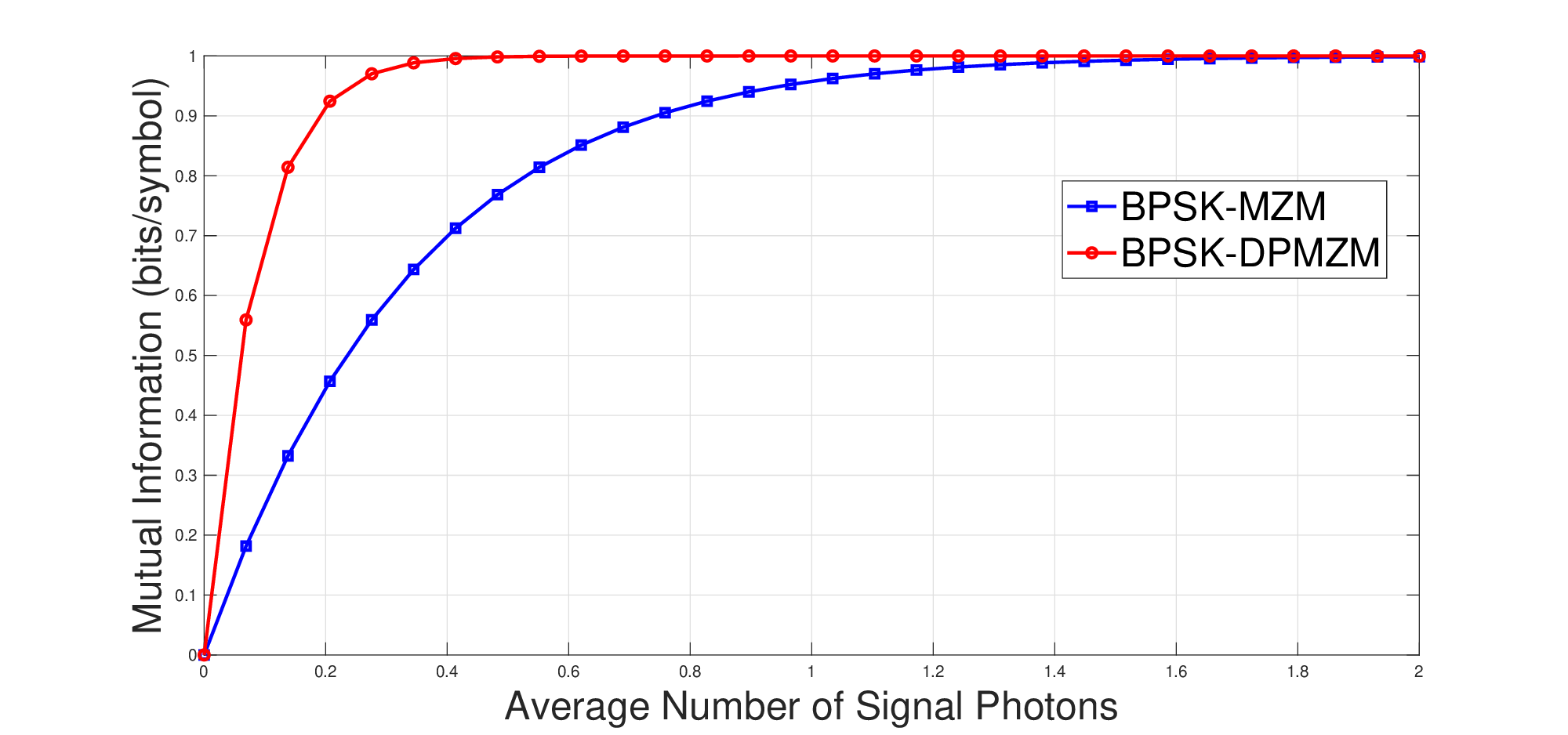}
\label{fig:fig3a}
}
\hfill
\subfloat[QPSK]{%
\includegraphics[width=0.5\textwidth]{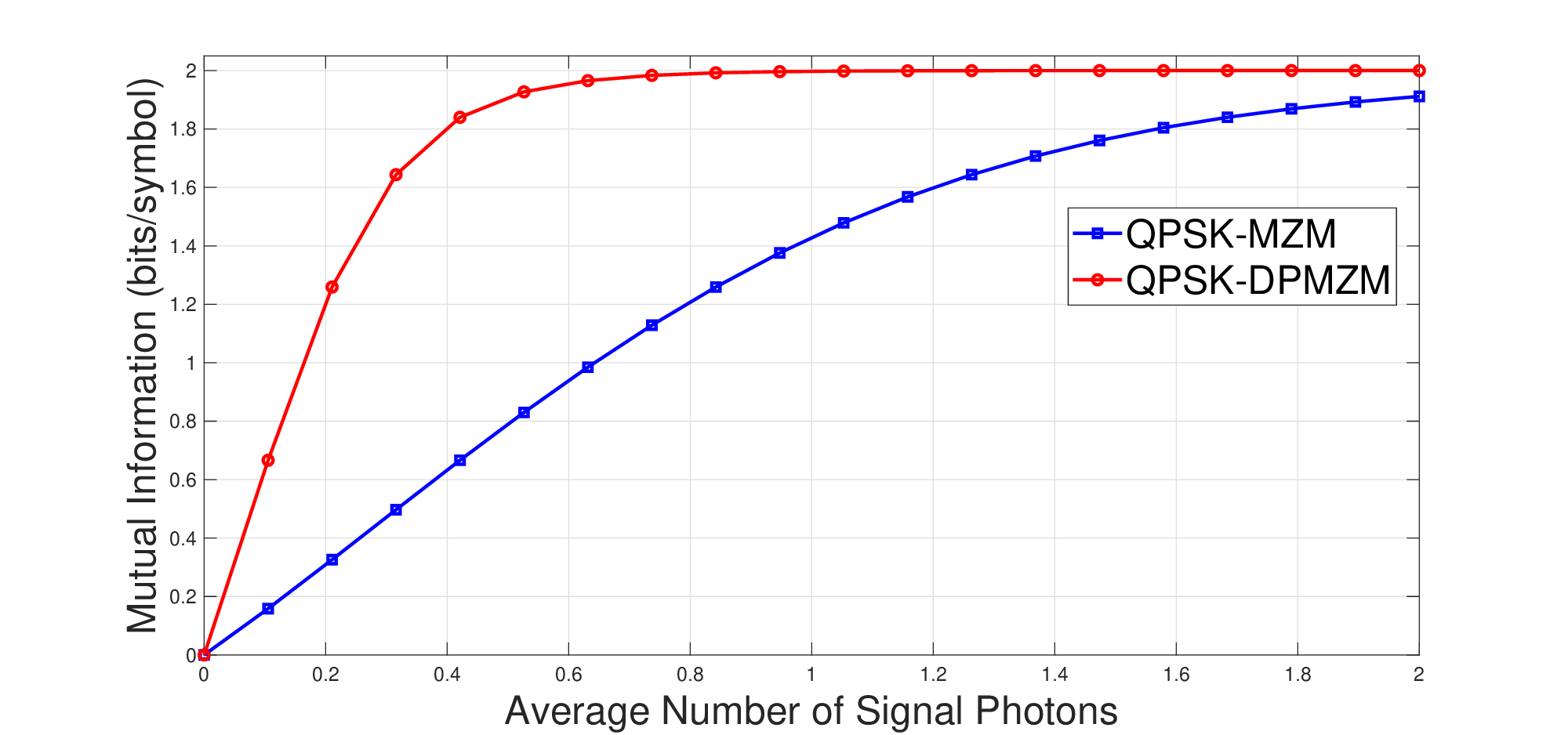}
\label{fig:fig3b}
}
\hfill
% \vspace{0.3cm}
\subfloat[16-QAM]{%
\includegraphics[width=0.5\textwidth]{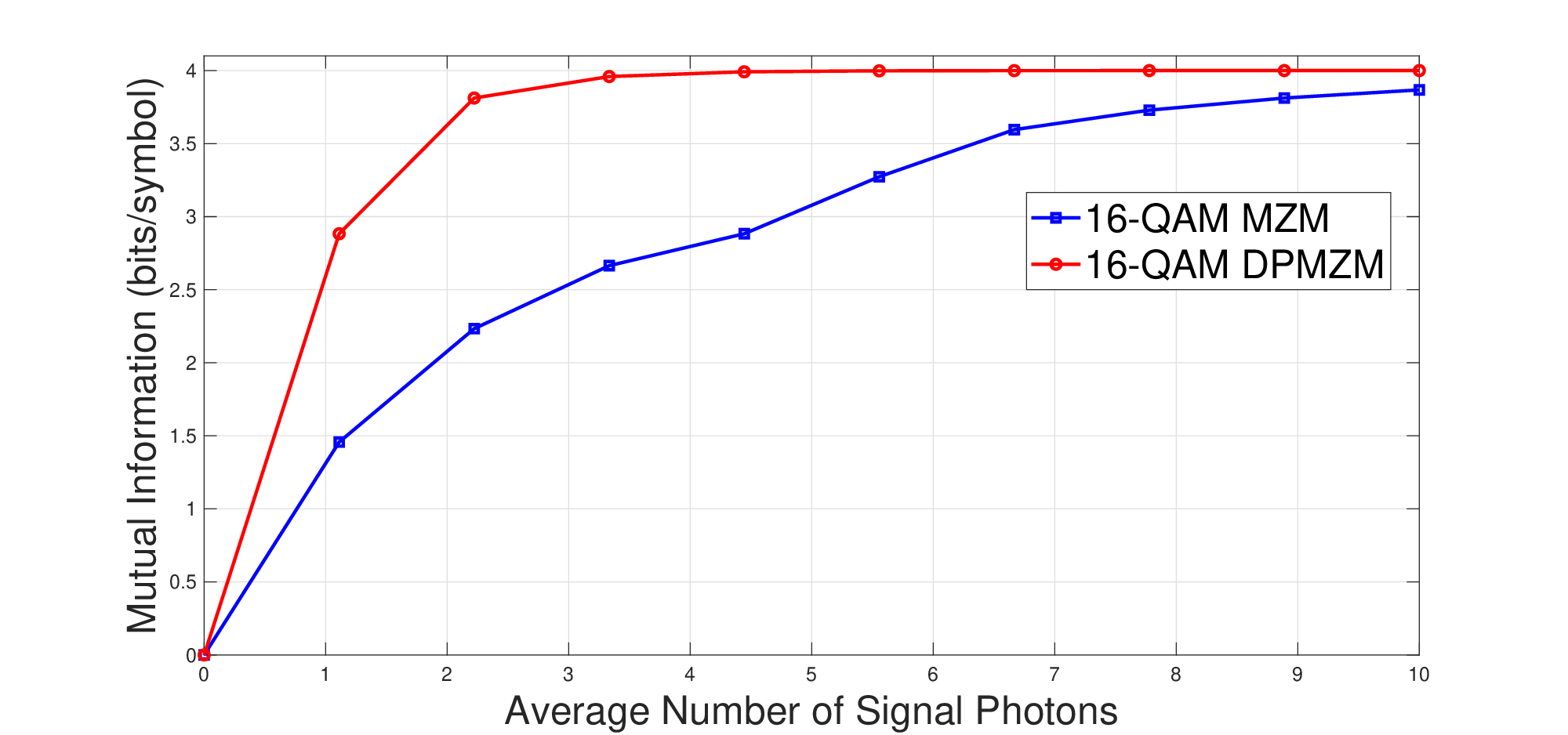}
\label{fig:fig3c}
}

\caption{Mutual Information performance versus average signal photon number for BPSK, QPSK, and 16-QAM schemes.}
\label{fig:fig3}

\end{figure}
Due to the dense constellation structure, the coherent states exhibit higher overlap, making detection more challenging compared to BPSK and QPSK. The classical 16-QAM receiver shows a slow decay in error probability with increasing photon number, requiring a significantly larger $N_s$ to achieve acceptable performance. As shown in Fig.~\ref{fig:fig2c} for 16-QAM modulation, at \(N_s = 16\), the proposed quantum-based DPMZM system achieves nearly eighth orders of BER improvement compared with quantum-based MZM and about fifth orders improvement compared with classical detection using DPMZM.  In contrast, the quantum receiver based on SRM demonstrates a much faster reduction in error probability. 
Fig. (\ref{fig:fig2a})-(\ref{fig:fig2c}) demonstrate that DPMZM-based quantum detection provides superior performance compared to classical detection for all modulation schemes, especially in low-photon conditions and higher-order modulation formats, making it highly suitable for NG-PON2 applications. 

Similarly, Fig. (\ref{fig:fig3a})-(\ref{fig:fig3c}) present the mutual information achieved under quantum detection schemes for BPSK, QPSK and 16-QAM modulation schemes. The results demonstrate that the quantum-based DPMZM-architecture attains higher mutual information at lower average photon levels compared with the quantum-based conventional MZM system, indicating enhanced photon efficiency and improved information transfer in the quantum detection regime. As the photon number increases, the DPMZM consistently converges more rapidly toward the theoretical mutual information limit. In particular, for 16-QAM, the DPMZM approaches the maximum achievable mutual information of 4 bits/symbol at significantly lower photon levels, highlighting its potential for high-capacity and photon-efficient quantum optical communication systems.

For practical high-speed implementation, the proposed quantum receiver can employ an InGaAs photodetector with 70--85\% quantum efficiency, 0.9--1.0~A/W responsivity, and 25--40~GHz electrical bandwidth, supported by a low-noise transimpedance amplifier. A 25--50~GBaud operation is feasible with appropriate coherent detection and digital signal processing. Practical limitations such as detector loss, electronic noise, phase noise, timing jitter, and finite ADC resolution may degrade the performance from the ideal Helstrom/SRM limits. These parameters provide a realistic basis for experimental validation of the proposed DPMZM-based quantum-enhanced NG-PON2 system.

\section{Conclusion} \label{conc}
In this work, we have proposed a DPMZM-based NG-PON2 communication system integrated with quantum modulation and detection techniques. We have also presented a detailed analysis of the SNR, BER, and mutual information of the DPMZM-based NG-PON2 quantum communication system. Our analysis revealed that a $6$ dB improvement in SNR can be achieved by using the DPMZM-based model compared to the MZM-based model. Furthermore, our numerical simulations demonstrate that the proposed quantum-based DPMZM system achieves orders-of-magnitude improvement in BER compared to the baseline quantum-based MZM system. The proposed system also demonstrated better mutual information performance across different adaptive modulation schemes, making it suitable for high-capacity and energy-efficient NG-PON2 and next-generation communication networks. Future research will focus on incorporating realistic receiver imperfections into the system model to better reflect practical deployment scenarios.

% In addition, the proposed quantum-based DPMZM system achieved nearly five orders of BER improvement for BPSK with \(N_s = 1\), nearly nine orders for QPSK with \(N_s = 5\), and nearly eight orders for 16-QAM with \(N_s = 16\), compared with the quantum-based MZM system.  Furthermore, our numerical simulation results show that the proposed quantum-based DPMZM system achieved orders of magnitude BER improvement compared with the quantum-based MZM system.

% \section*{Acknowledgment} 

% The work of Neel Kanth Kundu was supported in part by the DST INSPIRE Faculty Fellowship (Reg. No.: IFA22-ENG 344), ANRF Prime Minister Early Career Research Grant (ANRF/ECRG/2024/000324/ENS), and IIT Delhi's New Faculty Seed Grant (MI02855G).
\bibliographystyle{IEEEtran}
\bibliography{references}

@article{dalla2015optimality,
  title={{Optimality of square-root measurements in quantum state discrimination}},
  author={Dalla Pozza, Nicola and Pierobon, Gianfranco},
  journal={Physical Review A},
  volume={91},
  number={4},
  pages={042334},
  year={2015},
  publisher={APS}
}

@book{proakis2008digital,
  title={{Digital Communications}},
  author={Proakis, J.G. and Salehi, M.},
  isbn={9780071263788},
  lccn={2007036509},
  series={McGraw-Hill International Edition},
  url={https://books.google.co.in/books?id=ksh0GgAACAAJ},
  year={2008},
  publisher={McGraw-Hill}
}

@article{Nesset2015,
author = {Nesset, Derek},
year = {2015},
month = {03},
pages = {1-1},
title = {{NG-PON2 technology and standards}},
volume = {33},
journal = {Journal of Lightwave Technology},
doi = {10.1109/JLT.2015.2389115}
}

@ARTICLE{Kondamuri2024,
  author={Ramtej Kondamuri, Shri and Kanth Kundu, Neel and Ghanshyala, Amit},
  journal={IEEE Wireless Communications Letters}, 
  title={{Quantum Communication in Weakly Turbulent Channels: Error Analysis and Capacity Bounds}}, 
  year={2024},
  volume={13},
  number={12},
  pages={3658-3662},
  doi={10.1109/LWC.2024.3485533}}

@ARTICLE{Pravindra2023,
  author={Kumar, Pravindra and Sharma, Navneet and Gautam, Kumar},
  journal={IEEE Transactions on Consumer Electronics}, 
  title={{Performance Comparison of Classical and Quantum-Based NG-PON2 Communication System}}, 
  year={2023},
  volume={69},
  number={4},
  pages={1035-1044},
  doi={10.1109/TCE.2023.3295169}}

@article{patentCN104836625A,
  author       = {},
  title        = {{Dual-Parallel Mach-Zehnder Modulator (DPMZM) High-Linearity Microwave Photonic Link Implementation System}},
  howpublished = {Chinese Patent CN104836625A},
  year         = {2015},
  note         = {Published Aug. 12, 2015},
  url          = {https://patents.google.com/patent/CN104836625A/en}
}

@article{Singh2025,
  author    = {K. Singh and S. K. Arya and S. Singla and P. Berwal},
  title     = {{Performance Study of Microwave Photonic Links by Considering the Effect of Phase Shifters and Bias Conditions on Dual-Drive Dual Parallel Mach--Zehnder Modulator}},
  journal   = {Journal of Optical Communications},
  volume    = {46},
  number    = {3},
  pages      = {507--513},
  year      = {2025},
  doi       = {10.1515/joc-2024-0012}
}

@ARTICLE{9102229,
  author={Li, Ruwei and Sun, Xingwu and Yang, Dengcai},
  journal={IEEE Photonics Technology Letters}, 
  title={{A Novel Decoupling Bias Control Technique for Dual Parallel Mach-Zehnder Modulator}}, 
  year={2020},
  volume={32},
  number={13},
  pages={815-818},
  doi={10.1109/LPT.2020.2997998}}

@article{CatalaLahoz2025,
  author    = {Cristina Catal{\'a}-Lahoz and Moritz Merklein and Choon Kong Lai and Duk-Yong Choi and Stephen J. Madden and Benjamin J. Eggleton and Jos{\'e} Capmany},
  title     = {{On-Chip Stimulated Brillouin Scattering to Linearize a Dual-Parallel Mach--Zehnder Modulator}},
  journal   = {Journal of Lightwave Technology},
  volume    = {43},
  number    = {11},
  pages     = {5244--5251},
  year      = {2025},
  month     = jun,
  doi       = {10.1109/JLT.2025.3546728}
}

@article{Zhang2023,
  author    = {Peide Zhang and David Lowndes and Milan Stefko and Daniel Oi and John Rarity},
  title     = {{Modelling and Experimental Testing of an Optical Synchronisation Beacon Designed for High-Loss Satellite Quantum Communication}},
  journal   = {IET Quantum Communication},
  volume    = {5},
  number    = {1},
  pages     = {52--65},
  year      = {2023},
  doi       = {10.1049/qtc2.12071}
}

@article{Kim2011,
  author    = {S.-K. Kim and W. Liu and Q. Pei and L. R. Dalton and H. R. Fetterman},
  title     = {{Nonlinear Intermodulation Distortion Suppression in Coherent Analog Fiber Optic Link Using Electro-Optic Polymeric Dual Parallel Mach--Zehnder Modulator}},
  journal   = {Optics Express},
  volume    = {19},
  number    = {8},
  pages     = {7865--7871},
  year      = {2011},
  doi       = {10.1364/OE.19.007865}
}

@article{Torrijos2023,
  author    = {L. Torrijos-Mor{\'a}n and C. Catal{\'a}-Lahoz and D. P{\'e}rez-L{\'o}pez and L. Xu and W. Tianxiang and D. P{\'e}rez-Galacho},
  title     = {{Linearization of a Dual-Parallel Mach--Zehnder Modulator Using Optical Carrier Band Processing}},
  journal   = {Journal of Lightwave Technology},
  volume    = {41},
  number    = {10},
  pages     = {2969--2975},
  year      = {2023},
  doi       = {10.1109/JLT.2022.3229265}
}

@article{Feng2024,
  author    = {H. Feng and T. Ge and X. Guo and B. Wang and Y. Zhang and Z. Chen and C. Wang},
  title     = {{High-Linearity Dual-Parallel Mach--Zehnder Modulators in Thin-Film Lithium Niobate}},
  journal   = {Photonics},
  volume    = {11},
  number    = {10},
  pages     = {987},
  year      = {2024},
  doi       = {10.3390/photonics11100987}
}

@article{Takeoka2014,
  author    = {M. Takeoka and S. Guha},
  title     = {{Capacity of Optical Communication in Loss and Noise with General Quantum Gaussian Receivers}},
  journal   = {Physical Review A},
  volume    = {89},
  number    = {4},
  pages     = {042309},
  year      = {2014},
  doi       = {10.1103/PhysRevA.89.042309}
}

@article{Notarnicola2024,
  author    = {M. N. Notarnicola and S. Olivares},
  title     = {{Employing Weak-Field Homodyne Detection for Quantum Communications}},
  journal   = {IEEE Transactions on Communications},
  year      = {2024},
  note      = {Early Access},
  doi       = {10.48550/arXiv.2405.14310}
}

@article{Warke2024,
  author    = {A. Warke and others},
  title     = {{Photonic Quantum Receiver Attaining the Helstrom Bound}},
  journal   = {arXiv preprint arXiv:2410.21800},
  year      = {2024},
  url       = {https://arxiv.org/abs/2410.21800}
}

@article{Burenkov2021,
  author    = {I. A. Burenkov and M. V. Jabir and S. V. Polyakov},
  title     = {{Practical Quantum-Enhanced Receivers for Classical Communication}},
  journal   = {AVS Quantum Science},
  volume    = {3},
  number    = {2},
  pages     = {025301},
  year      = {2021},
  doi       = {10.1116/5.0042355}
}

@book{Gianfranco2015,
author = {Cariolaro, Gianfranco},
title = {{Quantum Communications}},
year = {2015},
isbn = {3319155997},
publisher = {Springer Publishing Company, Incorporated},
}

@article{Kato1999,
  author    = {Kentaro Kato and Masao Osaki and Masahide Sasaki and Osamu Hirota},
  title     = {{Quantum Detection and Mutual Information for QAM and PSK Signals}},
  journal   = {IEEE Transactions on Communications},
  volume     = {47},
  number     = {2},
  pages      = {248--254},
  year       = {1999},
  month      = feb,
  doi        = {10.1109/26.752519}
}
\end{document}